# MFB: A Mid-Frequency-Band Space Gravitational Wave Observer for the 2020 Decade

Peter F. Michelson[1], Robert L. Byer[1], Sasha Buchman[1],
Ilya Mandel[2], John Lipa[1], Shally Saraf[1]

[1]*Stanford University, Stanford, California 94305 United States*
[2]*Monash University, Wellington Rd, Clayton VIC 3800, Australia*

**Introduction:**

We make the case for the early development of a **M**id-**F**requency-**B**and (MFB) gravitational wave (GW) observatory in geosynchronous orbit (73,000 km arm), optimized for the frequency band 10 mHz to 1 Hz. MFB bridges the science acquisition frequencies between the ground observatories LIGO[1,2]/VIRGO[3] (4/3 km arm - as well as future planned ones 10/40 km arm), and the milli-hertz band of LISA[4] (2.5 Gm arm)- with useable sensitivity extending to 10 Hz. We argue that this band will enable the timely development of this game-changing field of astrophysics, with observations of medium mass Binary Black Holes (BBH) and Binary Neutron Stars (BNS) sources prior to their mergers in the LIGO frequency range as well as Extreme Mass Ratio Inspirals (EMRI)s and mergers of supermassive BBH within the main detection band. MFB is better placed than LISA to access this exciting frequency region.

A combination of high and low frequency GW observations from ground and space-based detectors is highly desirable to achieve the next key breakthroughs in our understanding of the new and dark Universe hinted at by electromagnetic wave astronomy. An MFB observatory builds on LISA technology as well as LIGO and adds significant new sources and science to GW astronomy in a timely way[5]. By reducing cost and taking advantage of already spent development costs for LISA, the MFB mission could be launched as much as a decade earlier.

**Key Science Goals and Objectives:**

The discovery of the abundance of BBH of tens of solar masses, the multi messenger astrophysics with BNS mergers[6] and the remarkable scientific results derived from the data solidify the case for GW astrophysics and astronomy. However, fundamental questions remain, including the lack of unification of General Relativity with the Standard Model (SM) and/or Grand Unified Theories[7], as well as the inconvenient fact that dark energy, dark matter and inflation, three fundamental elements of our cosmological model, are not part of the SM. MFB's mid-frequency band has attracted significant attention from the GW community[8,9], particularly as the detections by LIGO/VIRGO[10] give estimated event-rates for MFB of $10^3$-$10^6$ per year[11,12]. For a similar signal-noise ratio (SNR), these signals that require 5 years integration times for LISA are observable in a few months by MFB[13,14]. Below we summarize the expected and potential astrophysical and astronomical results from MFB observations in its 10 mHz to 1 Hz frequency band - new sources are also likely to be detected:

*Enhanced BBH parameter estimation:* MFB observations will occur well before these chirping signals enter the LIGO/VIRGO band, resulting in precise source parameter estimations, thereby allowing "coherent tracking" across the entire frequency band and resulting in precise tests of the "no hair theorem" by measuring the space-time multipoles, as well as other GR tests in the strong-gravity regime[9].

*Sky localization:* Binary neutron star GW sources will have quasi-constant frequencies for many years over most of the MFB, thus allowing the antenna to use the 2 AU diameter of the solar orbit as the baseline for sky localization to an estimated few arc-minutes[8]. Consequently, the host galaxies of the neutron-star binaries could be identified to distances of $r_{max} \lesssim 500$ Mpc, making possible the study of the environment of the binary well in advance of coalescence.

*Type IA supernova progenitors:* The question of the creation of type IA supernovae would be answered by an IA observation and the detection, or lack thereof, of a coincidental GW event[8].

*Mergers in the presence of third bodies*: Signals of GW mergers in the MFB will carry the imprint of any nearby third bodies, such as massive black holes or centers of massive core-collapsed globular clusters[8].

*Evolutionary history of compact object binaries*: The MFB observatory will expand the frequency spectrum coverage for BBH and Neutron star Black Hole (NBH) events, thus improving the understanding of the evolution and formation of these objects[8].

*Stochastic background*: Detections at MFB frequencies could possibly allow the observation of the cosmological GW background[15,16].

*Primordial Black Hole (PBH) formation:* PBHs are theorized to have been generated by various models[17,18,19,20] of the early Universe, resulting in a wide range of PBH masses - from the Plank mass to many orders of magnitude above $M_\odot$[21]. MFB's numerous GW detections, with accurate parameter estimation, could allow differentiation between these models.

*Element formation***:** Additional detections of NBH or BNS mergers will improve the understanding of the formation of heavy elements[22,23].

*Massive and Supermassive Black Holes:* MFB will characterize the parameters of coalescing BBHs with masses in the $10^3$-$10^8$ $M_\odot$ range[24], with precisions comparable or better than LISA - the larger amplitude modulation due to the diurnal rotation of the array should compensate for a lower SNR[25].

*Massive Black Hole Formation:* MFB will search for mergers leading to the creation of the massive black holes inhabiting the centers of galaxies, thus validating or negating different proposed scenarios for their formation[26]. For the MFB sensitivity band, the potential nuclei for mergers would be in the range of $10^2$-$10^3$ $M_\odot$ and would have been generated by first generation stars[8]. This would help establish the distribution of black hole seeds from population III stars and thus probe the formation of galactic structure.

*Intermediate Mass Black Holes (IMBH):* MFB has the optimal band for IMBH, ~$10^3$ $M_\odot$, detection; observable as either mergers or inspirals of compact stellar mass objects[7]. Under the assumption that IMBHs are central to globular clusters[27], observations of their mergers would help the understanding of the dynamics of the globular clusters[8,9].

*EMRI:* MFB will observe with good SNRs the spiraling of small black-holes (a few to 10 $M_\odot$) into larger ($10^2$ - $10^6$ $M_\odot$) holes; the Extreme and Intermediate Mass Ratio Inspiral binary systems[28]. These astrophysical objects are expected to radiate predominantly in the region of the GW band where LISA and MFB achieve their best sensitivities.

*Improved measurements of the Hubble constant, $H_0$* [29]*:* Detections GW170817[30] and GRB 170817A [31,32] give a 'GW Hubble constant'[33] $H_0^{GW} = 70.0\,^{+12.0}_{-8.0}\,\frac{\text{km}}{\text{s}\cdot\text{Mpc}}$. MFB's high event rates will allow (with ~100 mergers) a determination of $H_0^{GW}$ to ≤ 5% and over a much larger volume of space[34].

*Galactic Binary Calibrators:* MFB will also study known galactic binaries containing stellar-mass objects whose physical properties and sky locations have already been identified through optical observations (the so called "calibrators"). In relation to stellar-mass binary systems, in particular white dwarf binaries, it should be said that the hundreds of millions of such systems in our own galaxy, forming a "noise background" in the LISA data, will not degrade the MFB data because of its poorer sensitivity in the GW frequency band where this background radiates.

**Technical Overview:**

The MFB concept envisions a geocentric spacecraft formation with arm length between 73,000 km (*gLISA*[35] in geosynchronous orbit) and 666,000 km (*Lagrange*[36] at Earth-lunar 3, 4, 5

Lagrange points). MFB's concept and technology have been studied for the past ten years at Stanford University, the Jet Propulsion Laboratory, the National Institute for Space Research, and at Space Systems Loral (SSL), resulting in a 2020 decade launch date, while using a conventional program development plan at a cost comparable to medium scale observatories launched by NASA in the previous decade. Similar to LISA, MFB will exchange coherent laser beams along its three arms[37] and synthesize interferometric combinations that are highly sensitive to gravitational radiation by applying Time-Delay Interferometry (TDI)[35,38,39] to its heterodyne measurements[40]. Figure 1 shows the characteristic strain sensitivities of MFB in the geosynchronous orbit compared to advanced LIGO and LISA, the early GW sources detected by advanced LIGO, and examples of expected sources for MFB and LISA; note that the characteristic strain $h_c(f)$ and the strain sensitivity $h(f)$ are related by $h_c(f) \equiv h(f)\sqrt{f}$.

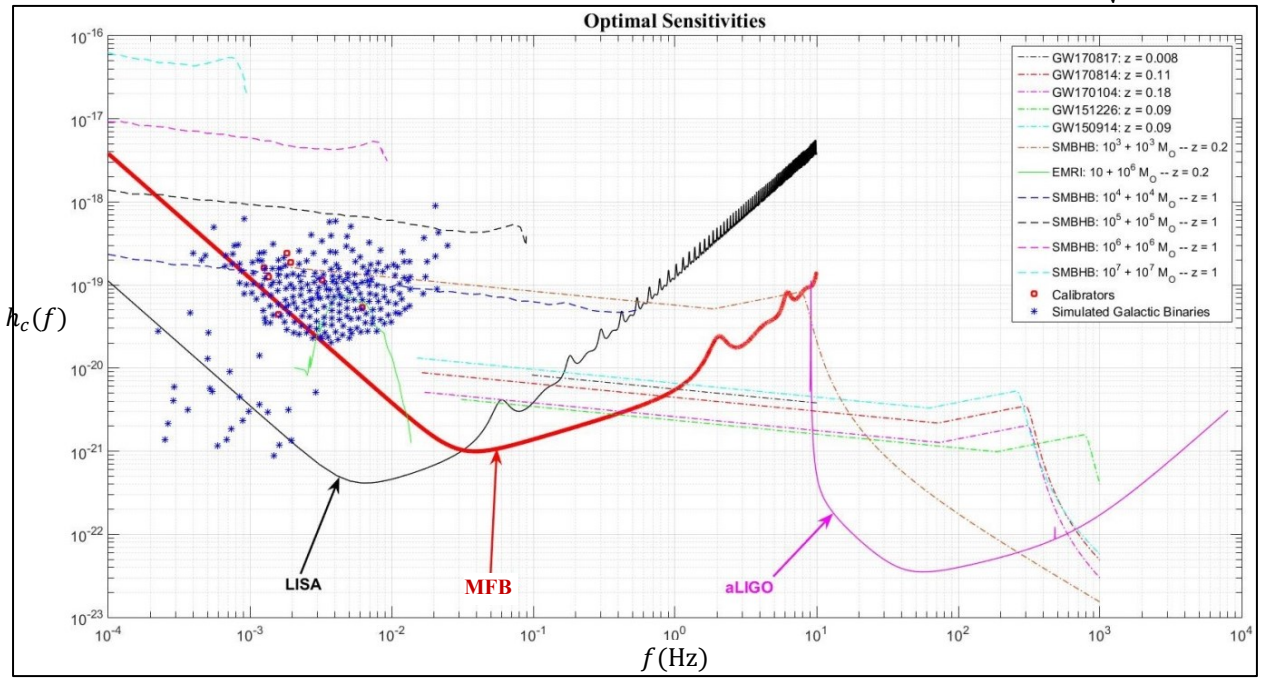

Figure 1. The characteristic strain $h_c(f)$ for MFB, LISA and aLIGO averaged over sources randomly distributed over the sky and the polarization states. Each of these curves is uniquely determined by the TDI *A*, *E*, and *T* data combinations associated with the data from each mission (see ref 41). For completeness we have included the amplitude of the early events detected by aLIGO, as a function of the Fourier frequency *f*, and other potential sources.

The MFB mission is based on three principles: 1) operate an instrument with a noise performance similar to LISA but with shorter arm-lengths (73,000 km to 666,000 km), resulting in maximum sensitivity in the 10 mHz to 1 Hz frequency range; 2) keep costs down by developing and testing critical technologies in parallel, utilizing small satellites, and 3) engage the international community in the effort to develop the instrument and mission in a community-wide science program. We show that the MFB mission is implementable at a cost of between $500 million and $1 billion and can be flown in the 2020's using a combination of parallel developments and significantly reduced complexity. Very important, compared with LIGO, is that source detections can routinely be made long before coalescence, greatly improving the options for accompanying electromagnetic observations. A technical challenge is the required space interferometry sensitivity of below 1 pm for arm lengths below about 200,000 km; where thermal and optical path length errors become dominant over photon noise.

Modern GW detectors are based on the measurement of the modulation of space-time caused by the passing of a GW between two or more 'free floating' test masses (TMs). In 1971 R. Weiss first promoted the concept of laser interferometry as the best method to achieve the precision required for the detection of GW[1]. Interferometers operating at the quantum noise limit are now the instrument of choice for ground-based detectors. The first LISA-like space-based detector was proposed in 1981[42,43] and similarly based on laser interferometry. Conceptually, a space GW detector is a modified Michelson interferometer consisting of three TMs and a ruler based on light. The difficulty arises due to the 'weakness' of the gravitational interaction and the 'stiffness' of space. GW amplitudes, known as 'strain amplitude $h$' and defined by $h \equiv dl / l$, are typically expected to be of the order $h \approx 10^{-20}$ for sources detectable by ground-based GW detectors. Measuring the strain of space-time to the necessary precision is equivalent to measuring the displacement of an atom at the distance of the Sun or 1/1000 the diameter of a proton in the 4-km path length of LIGO.

Because its arm length is about a factor of between 4 and 30 shorter than that of the LISA mission[44,45], the MFB mission will have optimum sensitivity to the GW spectrum that is between that accessible by LISA and that of ground-based interferometers[46,47]. The MFB mission will complement the scientific capabilities of both LISA and ground-based interferometers and meet the GW science objectives stated in the NASA's Astrophysics Visionary Roadmap[48] and Science Plan[49] documents in a much earlier time frame than LISA.

**Mission Architecture, Performance and Cost**

The orbit chosen for the MFB mission is geosynchronous (with an option for a geocentric 666,000 km arm-length orbit with spacecraft in the Earth-Lunar 3, 4, and 5 Lagrange points) as opposed to the more common geostationary version used for commercial broadcasting and previously proposed for GW missions[50,51] as alternatives to the LISA mission in response to the NASA's Request for Information # NNH11ZDA019L[52]. Geostationary orbits have nearly zero inclination, eccentricity, and east-west drift to allow them to "hover" directly above the same equatorial longitude and appear fixed in the sky to a ground user, which simplifies ground terminal designs. Geosynchronous orbits have a small inclination and allow modest daily motion to occur as long as it repeats over 24 h. A GW mission can allow inclination to vary as this offers advantages in source identification and to the satellite design to minimize complexity and cost. Three satellites in geostationary or geosynchronous orbit form an equilateral triangle of approximately 73,000 km arm-length. Our mission concept could rely on a single, dedicated launch or three shared launches. In the single-launch option the launcher enters a Geosynchronous Transfer Orbit (GTO), and from there it is transferred to an inclination drift orbit from which the three satellites are deployed to their final locations. Sizing of the satellites is driven by the main requirements for the GW payload. Payload weight, power and size is similar to that of LISA, LPF (125 kg plus telescopes)[53] and other GW missions studied previously. Satellite and propellant weight are estimated at 300 Kg and 50 Kg respectively (tables 1 and 2).

**Table 1. Satellite hosting requirements for MFB payload.**

| | |
|---|---|
| Mass | 200 kg |
| Power | < 500 W |
| Size | 1x1x0.5 m |
| Thermal | 10 – 30 °C |
| Data rate | 10 Mbps peak |

**Table 2. MFB satellite mass breakdown**

| | Bus (kg) | Payload (kg) | Propellant (kg GN2) |
|---|---|---|---|
| Satellite | 300 | 200 | 50 |
| Constellation | 900 | 600 | 150 |
| Total: 1650 kg | | | |

Using a Falcon 9 launch provides significant cost savings compared to alternatives such as the Evolved Expendable Launch Vehicle (EELV) program or Ariane launch costs. Proper selection of mission inclination reduces system cost by minimizing the on-orbit satellite propulsion requirements for station-keeping. A technique known as "inclination drifting" is used to allow satellite to maintain an inclination range without use of any significant propellant. For our GW mission concept, selection of a Right Ascension of the Ascending Node (RAAN) in the 253° to 330° range allows inclination to start at and stay below 3° for at least 5 years, figure 2. After GTO insertion, the launch vehicle carries the satellites during the orbit circularization phase and then drifts around the orbit to release them roughly 120° apart. This process takes approximate 2 months. At the target inclination and RAAN, all three satellites will be operated in the same orbit plane, and they will stay in a stable relative formation with minimal propulsion activity.

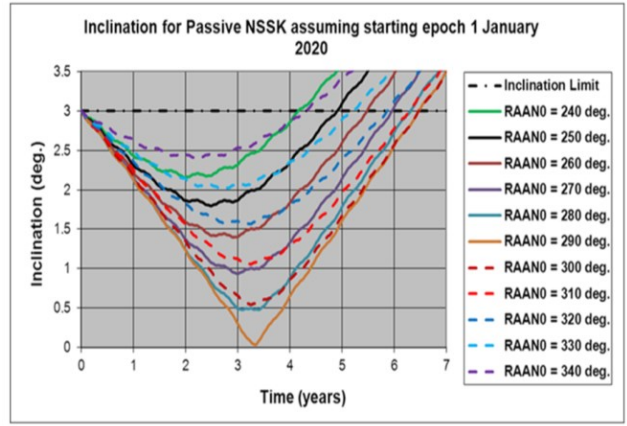

Figure 2. MFB trajectory inclination angle as function of time and value of the RANN angle, for starting date of January 1, 2020.

Two configurations of the satellites are considered; a concept by SSL of mothership with two daughter ships (figure 3) and the standard three identical spacecraft. A summary of the JPL estimated launch margins for the two configurations, Solar Electric Propulsion (SEP) and chemical propulsion, and Atlas /Falcon launch vehicles is shown in table 3.

Since ground operations can be a significant portion of total mission costs, the dedicated system uses cross-links to allow one satellite to be the central gateway and mission control center that coordinates all activities through it. Heritage Ka-band systems with 0:75 m reflectors exist, providing up to 10 Mbps peak data rate. All routine operations have telemetry and control and mission data to fixed ground control centers. Orbit determination may be done by infrequent ranging, onboard GPS, or cross-link microwave ranging, while the time coordination is controlled by the main satellite.

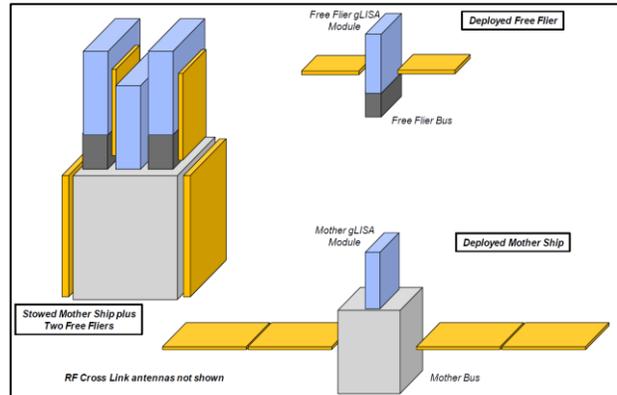

Figure 3. Mothership + 2 Daughterships Concept.

**Table 3. Launch margins into geosynchronous orbit.**

| Concept | Design | Propulsion | Δv(m/s) | Vehicle | LV Allocation | Margin |
|---|---|---|---|---|---|---|
| Geosynchronous | Mothership + 2 Daughterships | Chemical | 1900 (M) / 100 (D) | Atlas V 431 | 7105 | 29% |
| | Mothership + 2 Daughterships (SSL Concept) | SEP | 1900 (M) / 100 (D) | Falcon 9 | 5755 | 66% |
| | 3 Sisterships | Chemical | 1800 | Falcon 9 | 5755 | 31% |
| | 3 Sisterships | SEP | 1800 | Falcon 9 | 5755 | 68% |

Science data return is maximized with more than 90 percent collection times daily except for eclipse seasons at the equinoxes. To avoid costly design features for thermal accommodations enabling science data acquisition during eclipses, the payloads can be shut down for 45 days each spring and fall. Hence, a 4 year on-orbit period will yield 3 years of science data. However, ultra-narrow bandpass filters developed over the last few years (0.1 – 4 nm at the laser frequency)[54,55,56,57] applied to the front windows of the telescopes would allow all-year operation. Science will be maximized by coordinating all activities on the three satellites to have synchronized maneuver times and housekeeping periods.

*Mission Performance*

The MFB sensitivity shown in Figure 1 assumes a residual acceleration noise in each spacecraft equal to $\sqrt{S_a(f)}$ =3.0×10$^{-15}$ m·s$^{-2}$Hz$^{-1/2}$ (with the noises defining it referred to as *low-frequency noises*), and a residual position noise: $\sqrt{S_L(f)} = 0.5$ pm·Hz$^{-1/2}$ (with the noises defining it referred to as *high-frequency noises*). Although these requirements might appear challenging, we believe there are no major roadblocks along the path to achieve them.

The laser interferometry configuration can be very similar to that of LISA[44,45]. On each spacecraft there are three optical modules: the laser frequency stabilization module, the phase-locking module, and the heterodyne interferometer. The laser frequency stabilization module consists of a master laser and a frequency stabilization control system[58]. The phase locking module consists of a slave laser and an offset phase lock loop with which the slave laser is phase-locked to the master. The heterodyne interferometer consists of two interferometer optical systems and their corresponding laser beam pointing control systems (if deemed to be necessary for the MFB trajectory). Both are bonded on a single piece of ultra-low expansion (ULE) glass baseplate to form a quasi-monolithic optical bench.

Two one-way heterodyne Doppler measurements are performed along each of the three arms by relying on two inertial sensing systems, one in each spacecraft, to measure the displacement between pairs of almost freely floating TMs. The distance between the TM and the interferometer optical bench is measured by a local short-arm interferometer on each spacecraft, while the displacement between the two interferometer optical benches is measured by the inter-spacecraft long-arm interferometer.

For the disturbance reduction system, there are two possible designs that could be adopted. One design relies on a single spherical TM of the type developed at Stanford University with optical readouts in the inertial sensor[59], while the other, more complex design would use a pair of cubic TM sensors similar to those tested in the LISA Pathfinder mission[60].

*Trajectory*

The trajectory of a satellite is determined by the influence of both gravitational and non-gravitational forces. In the case of the MFB satellites the dominant factors determining their trajectories are the gravitational field of the extended Earth and the Moon, the monopole gravitational field of the Sun, and solar radiation pressure. These forces will result in variations of the array's arm lengths and the triangle's enclosed angles. Distance variations will produce Doppler frequency shifts of the received laser beams that will require use of an onboard microwave frequency reference for removing the frequency offset from the heterodyne measurements.

Changes in the subtended angles instead could in principle require a pointing control mechanism to align the onboard optical telescopes. In an orbit analysis done for an array in a geostationary trajectory[37] (fairly well representing the geosynchronous case discussed in this paper) the inter-spacecraft relative velocities are periodic functions of period equal to 24 hours and amplitude less than 0.7 m/s (see figures 4,5,6). This general behavior remains largely the same during a nominal mission duration of five years.

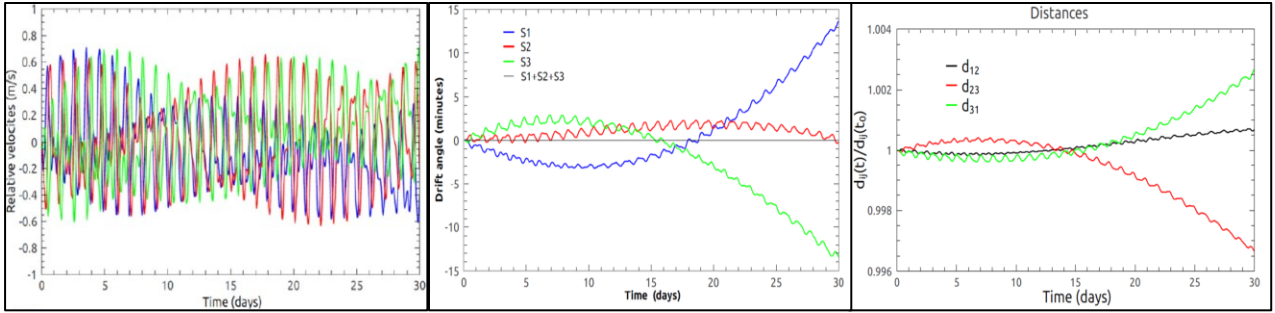

**Figure 4.** Time-dependence of inter-spacecraft velocities.

**Figure 5.** Drift angles between the spacecraft.

**Figure 6.** Time dependence of distance between the spacecraft.

The relative velocities will induce Doppler shifts in the laser frequencies that will have to be removed from the heterodyne measurements. This can be done by either relying on an onboard Ultra Stable Oscillator (USO)[61] or by generating the needed microwave signal with an onboard optical-frequency comb coherent to the frequency of the onboard laser[62]. The former has been extensively studied for the LISA mission, and it has recently been shown to introduce additional noise correlations in the resulting TDI combinations[63] due to the use of sidebands for calibrating out of the TDI measurements of the USO noise. The latter would entirely avoid this increased-noise effect due to the calibration procedure as it would not require use of sideband modulations[50].

In order to adopt this alternative heterodyne measurement technique, however, modifications of the interferometric design already studied for the LISA mission (and upon which MFB will mostly rely on) might be required. An optical frequency comb subsystem designed to meet the MFB requirements[64,65] is being considered as a design enhancement. The time-variation of the angles enclosed by the triangular constellation are plotted in figure 5. The values shown correspond to the differences between each angle's value at time $t$ and the 60° value at time $t = 0$. During the first two weeks, i.e. the time between two consecutive station-keeping maneuvers, the enclosed angles do not change much, remaining within the ±3 arc-minute range. Depending on the size of the adopted optical telescopes, the angles variations might need to be corrected by using laser beam pointing control with a fast-steering mirror. Figure 6 shows the time dependence of the distance between the three MFB satellites.

GW detection windows would extend to two-week periods due to 2 h east-west orbit trim; while maintaining the relative velocities between satellites to less than ± 0.5 m/s. The variations of the angles between the arms of the constellations are less than ± 3 arcmin, and the fractional arm length variations less than $5\times10^{-4}$. Large orbit formation adjustments could be done during the spring or fall eclipse seasons while the science payload is off.

Sensitivity normal to the ecliptic plane is less than that of LISA due to the reduced out-of-plane motion of the observatory. However, higher gravitational-wave harmonics detectable due to the shorter baseline of the MFB constellation provide a significant improvement in the position determination of Massive BBHs (MBBH). Locating spinning black holes in a MBBH is much more accurate than would be expected from the modulation produced by the precessing of the LISA plane alone[66].

In 2011 several alternatives were investigated by NASA[67,68] As part of that study, an international collaboration proposed a mission called LAGRANGE (Laser Gravitational- wave ANtenna in GEocentric orbit)[36] with an orbit at the L3, L4, and L5 Lagrange points of the Earth-Moon system, and with 666,000 km lengths arms. The geocentric orbit coupled with a single spherical TM per spacecraft with an acceleration noise of less than $3\times10^{-15}$ m·sec$^{-2}$Hz$^{-1/2}$ and a measurement precision of 8 pm Hz$^{-1/2}$ led to a strain sensitivity of $10^{-19}$ at 0.01 Hz.

**Technology Drivers:**
- Recent developments in the aerospace industry, designed to satisfy the growing demands for low-cost satellites and launch vehicles, have opened up a broader set of opportunities for shorter development cycle and lower cost GW missions. With this background, in 2013 we started to explore the scientific, technical, programmatic, and cost advantages of flying a geocentric GW mission with off-the-shelf satellites. Today, by relying on technologies existing in Europe and the United States, and the possibility of adopting the European drag-free system recently demonstrated by the LISA Pathfinder mission, we calculate that the MFB mission will reach its best strain sensitivity at a level close to that of LISA but over a frequency region that is higher by roughly a factor of 100, i.e. from about 10 mHz to 1 Hz; see Fig. 1 for a 73,000 km arm length orbit.

*The Modular Gravitational Reference Sensor - MGRS*
- A critical component for the MFB mission is the inertial reference system or Modular Gravitational Reference Sensor (MGRS). We plan to exchange the LISA drag-free sensors for a US-developed and flight proven design, leading to improved resolution and reliability. The electrostatically forced gravitational reference cube pairs of test masses of LISA will be replaced with single unsupported spherical test masses and the gaps to the housing will be increased by about an order of magnitude. The capacitive sensors and electrostatic forcing will be dispensed with and an all-optical readout system will be used instead. The sphere will be spun at ~10 Hz with its axis normal to the interferometer plane and the polhode motion damped magnetically prior to observation runs. This concept is based on the Gravity Probe B (GP-B) drag-free system[69] flown in 2004 and has been under development since 2005.
- A full-scale prototype model with an acceleration noise performance requirement of $10^{-14}$ m·sec$^{-2}$Hz$^{-1/2}$ and a goal of $3\times10^{-15}$ m·sec$^{-2}$Hz$^{-1/2}$ at frequencies between 0.1 mHz and 1 Hz, is under development. At the minimum requirement of $10^{-14}$ m·sec$^{-2}$Hz$^{-1/2}$ for the sensor, a space-based gravitational wave antenna will detect tens of massive black hole mergers per year and hundreds of galactic binary sources. The more challenging performance goal matches the LISA mission requirements level[70].
- The MGRS will use a spherical TM 7 cm in diameter, inside a housing where the gap to the walls equals the ball radius. The TM location in the housing will be determined to the nanometer level by an eight-beam differential optical shadow sensor (DOSS)[71]. The principal advantages of this design are[72] a) no active forces are applied to the TM, b) large gaps to the housing reduce patch effect forces dramatically[73], c) simplicity and reliability, d) minimization of ancillary electronics and control loops and e) long flight heritage of the central technology[74,75,76]. We have developed analytical models to demonstrate the enhanced performance of the full-scale design[77] and have completed the scaled down (1/3 version) laboratory prototype instrument[78]. An interferometer system is used to monitor the TM surface to the picometer level.
- A precision charge management system for the MGRS, using compact and low-power UV LEDs, has been developed and successfully flown on a small-sat in 2014 by Stanford University in collaboration with NASA Ames[79].
- In a gravitational wave antenna configuration, the MGRS will allow the detection of low frequency sources of gravitational radiation with simpler technology and at much lower cost than LISA. Its implementation would lead to a less constrained budget, allowing a much earlier launch than is likely for LISA. Further, our MGRS design can be shown to have better performance than the present LISA approach using cubes, allowing a cheaper, better path to a flight mission. Our analysis shows that this design is capable of exceeding the LISA reference sensor specification of $3\times10^{-15}$ m·sec$^{-2}$Hz$^{-1/2}$ at frequencies between 0.1 mHz and 10 Hz[77].

**Cost Estimates:**

Three main factors contribute to the performance and cost of a space detector: 1) the orbit of the three-spacecraft constellation, reflected in the cost of the launch vehicle 2) the design of the drag-free TMs contained within the satellites, reflected in the cost of payload and flight systems and 3) the laser interferometer measurement system operating between $\sim 10^5$ km and $\sim 10^6$ km to picometer precision, reflected in the cost of payload and flight systems. A 2016 Team-A JPL study has produced cost estimates for a geosynchronous/geocentric orbit mission with an estimated $150M payload cost, for the cases of NASA and Surrey spacecraft buses, a Falcon 9 launch vehicle and +/- 30% off nominal costs; see table 4. Reserves are at 30% excluding launch vehicle and numbers need to be increased by 6.7% for 2016 to 2019 inflation. The nominal costs $MFY19 with Surrey and NASA buses are then $760M and $890M, while the cost of mission with the more expensive NASA bus and an additional 30% above reserves is $1160M.

**Table 4. Cost estimate for MFB with:**

| Costs $M FY16 | Surrey bus | | | NASA bus | | |
|---|---|---|---|---|---|---|
| | -30% | Nominal | +30% | -30% | Nominal | +30% |
| WBS 1,2,3 Proj Mgmt, Proj SE, MA | $30M | $40M | $50M | $40M | $50M | $70M |
| WBS 4 Science | $30M | $40M | $50M | $30M | $40M | $50M |
| WBS 5 Payload | $110M | $150M | $200M | $110M | $150M | $200M |
| WBS 6 Flight System | $110M | $160M | $210M | $160M | $230M | $300M |
| WBS 7 and 9 MOS/GDS | $40M | $60M | $80M | $40M | $60M | $80M |
| WBS 10 ATLO | $0M | $0M | $0M | $0M | $0M | $0M |
| WBS 11 EPO | $0M | $0M | $0M | $0M | $0M | $0M |
| WBS 12 Mission Design | $0M | $0M | $0M | $10M | $10M | $10M |
| Reserves | $90M | $130M | $170M | $110M | $160M | $210M |
| Launch Vehicle | $90M | $130M | $170M | $90M | $130M | $170M |
| **TOTAL PROJECT COST** | $500M | $710M | $930M | $590M | $830M | $1090M |

For comparison, LISA, the benchmark space detector, has been studied and developed since 1993, and was budgeted in 2012 at $2.1 billion dollars[80].

Presently, ESA has approved a 2.5 Gm LISA at a cost of about $2.5 billion[81], to be flown 'not before' 2034. We here propose MFB, with a launch time in the 2020's that is both a challenge and a promise to the community.

The option of relying on off-the-shelf satellites in a geocentric orbit, which was never considered by previous GW mission concept studies[82], will result in a GW mission cost compatible with that of an astrophysics probe-class mission. Our estimate is based on the conclusions reached by a NASA Architecture Team (A-TEAM) study performed at the Jet Propulsion Laboratory in late January 2016 that looked at various alternatives to LISA. Such a study relied on previous TEAM-X cost estimates of other GW mission concepts[67] updated with newly available satellites and launching vehicle costs, the option of using a single TM onboard each satellite[72,83], and reduced costs for the constellation to communicate to the ground.

**Schedule:**

The MFB program would start with a 3-4-year campaign of two critical technology demonstrations in parallel on small satellites. The first technology development flight would verify the advanced drag-free technology for the TM and cost ~ $10M. The second would deploy a dual-satellite laser ranging system with interferometry measurements. We expect this mission to cost ~ $25M for each satellite and mission ops. A core team would design and build the well-known satellite and science instrument components[84,85]: satellites, telescopes and standard electronics. A review would establish the TRL of the critical technologies (currently estimated at ~TRL 4) and the program would proceed as appropriate with building the instruments over 3-4 years following the establishment of high TRLs via the small satellite missions. The observation program would be designed to extend over five years with data analysis continuing for at least 3 years post-mission.

**Organization, Partnerships, and Current Status:**

It is envisioned that if funded the MFB mission would include a large range of participating organizations, essentially involving the entire LISA and LIGO communities and other interested parties. Optimally the program would be directed by a team of academic scientists, on the model of LIGO or Fermi.

**Conclusions**

a) A mid-band GW detector will achieve the most important science goals of LISA listed in the 2010 astrophysics decadal survey, "New Worlds, New Horizons"[86].
b) Measurements of black hole mass and spin from massive BHB will be important for understanding the significance of mergers in the building of galaxies.
c) An equally powerful test will be provided by the mergers of massive BHB by comparing actual GW forms to the highly detailed numerical simulations performed by modern general relativistic hydrodynamics codes with dynamical space-time evolution[87].
d) Potential for discovery of waves from unanticipated or exotic sources, such as backgrounds produced during the earliest moments of the universe, dark energy signals or cusps associated with cosmic strings.
e) MFB observations will complement the scientific capabilities of both LISA and LIGO/VIRGO and meet the GW science objectives stated in the NASA's Astrophysics Visionary Roadmap[48] and Science Plan[49].
f) Geocentric orbits present decisive advantages over heliocentric ones by reducing the launch weight by half and increasing the telemetry and command bandwidth capability by more than two orders of magnitude. Augmented requirements for thermal control and Doppler shift compensation are well within the present technology capabilities of active thermal control multi-layer insulation and phasemeters.
g) A single spherical TM per spacecraft – with a long flight heritage - further decreases the complexity and weight of the experiment. Note, that alternate TM and interferometry designs can replace the proposed ones if these systems have reached high TRL and are cost effective.
h) The MFB can be developed and deployed in 7-10 years, well in advance and at less than half the cost of ESA's proposed LISA mission, while also providing a technical pathfinder for LISA.

**REFERENCES**

[1] R. Weiss, Quarterly Progress Report of RLE, MIT **105,** 54 (1972)
[2] LIGO Caltech
[3] VIRGO Pisa
[4] O. Jennrich, *et al.* NGO assessment study report (Yellow Book) 〈hal-00730260〉 (2012)
[5] P.F. Michelson ASTRO 2020 Science White Paper (2019)
[6] B. P. Abbott, *et al.* Astrophys. J. Lett. **848(2),** L12 (2017)
[7] C. M. Will, Living Rev. Relativity, **9** (2006)
[8] I. Mandel, A. Sesana and A. Vecchio, Classical Quant. Grav. **35(5)**, 054004 (2018)
[9] P.W. Graham *et al*. arxiv: 1711.02225v1 (2017)
[10] B. P. Abbott *et al.* arXiv:1811.12907v2. (2018)
[11] K. Belczynski, *et al.* Astrophysical J. **819(2)**, 108 (2016)
[12] K. Belczynski *et al.* Nature, **534**, 512 (2016)
[13] A. Sesana, Phys. Rev. Lett. **116**, 231102 (2016)
[14] M. Tinto, and J.C.N de Araujo, Phys. Rev. D, **94**, 081101(R) (2016)
[15] V. Mandic *et al*. Phys. Rev. Lett, **109**, 171102 (2012)
[16] T. Callister *et al*., Phys. Rev. X, **6**, 031018 (2016)
[17] S. Clesse and J. Garcìa-Bellido, Phys. Rev. D **92**, 023524 (2015)
[18] K. Jedamzik and J.C. Niemeyer, Phys. Rev. D **59**, 124014 (1999)
[19] J. Garriga and A. Vilenkin, Phys. Rev. D **47**, 3265 (1993)
[20] J. Garriga, A. Vilenkin and J. Zhang, J. Cosmol. Astropart. Phys. **02**, 064 (2016)
[21] H. Deng and A. Vilenkin, J. Cosmol. Astropart. Phys. **12,** 044 (2017)
[22] D. Kasen *et al*. Nature, **551**, 80 (2017)
[23] F.-K. Thielemann *et al.* Annu. Rev. Nucl. Part. S. **67**, 253 (2017)
[24] K. Belczynski *et al*. Astrophysical J. **789(2)**, 120 (2014)
[25] S.T. McWilliams "Optimal orbits for eLISA science", 10th LISA Symposium Florida, (2014).
[26] M. Volonteri, Astron. Astrophys. Rev. **18(3),** 279 (2010)
[27] C-J. Haster *et al.* Astrophysical J. **832(2),** 192 (2016)
[28] P. Amaro-Seoane *et al.* Classical Quant. Grav. **29(12)**, 124016 (2012)
[29] The LIGO Scientific Collaboration *et al*. Nature **551**, 85 (2017)
[30] B. P. Abbott *et al*. Phys. Rev. Lett. **119**,161101 (2017)
[31] A. Goldstein, *et al.* Astrophys. J. Lett. **848(2),** L14 (2017)
[32] B. P. Abbott, *et al.* Astrophys. J. Lett. **848(2),** L13 (2017)
[33] B.F. Schutz, Nature **323**, 310 (1986)
[34] W. Del Pozzo, Phys. Rev. D, **86**, 043011 (2012)
[35] M. Tinto, *et al*. Rev. Sci. Instrum. **86**, 014501, (2015)
[36] J W Conklin, *et al.* arXiv:1111.5264v2 (2011)
[37] M. Tinto, *et al*. Classical Quant. Grav. **32**, 185017 (2015)
[38] M. Tinto *et al*. Astroparticle Physics, **48**, 50 (2013)
[39] J. W. Armstrong, F. B. Estabrook, M. Tinto, Astrophys. J. **527(2),** 814 (1999)
[40] M. Tinto and S.V. Dhurandhar, Living Rev. Relativity, **17**, 6 (2014)
[41] T.A. Prince *et al.* Phys. Rev. D, **66**, 122002 (2002)
[42] J.E. Faller and P.L. Bender, *Abstract* for the Second International Conference on Precision Measurements and Fundamental Constants, Gaithersburg, MD, 8-12 June (1981)
[43] J.E. Faller and P.L. Bender, *A Possible Laser Gravitational-Wave Experiment in Space*, Precision Measurements and Fundamental Constants II, NBS Special Publication 617 (U.S. Govt. Printing Office, Washington, D.C., 689-690 (1984)
[44] P. Bender *et al*. Pre-Phase A Report, MPQ233 (1998)
[45] LISA: "Unveiling a hidden Universe", ESA publication # ESA/SRE (2011)
[46] LIGO: Caltech
[47] VIRGO: Pisa

[48] NASA Astrophysics Visionary Roadmap (2013)
[49] NASA Science Plan (2014)
[50] M. Tinto *et al.* arXiv:1111.2576 (2011)
[51] S. T. McWilliams arXiv:1111.3708v1 (2011)
[52] NASA solicitation # NNH11ZDA019L: "Concepts for the NASA gravitational-wave mission."
[53] LPF: (LISA Pathfinder) Mission (2019)
[54] J. Floriot, F. Lemarchand, M. Lequime, Optics Communications **222**, 101 (2003)
[55] J. Floriot, F. Lemarchand, M. Lequime, Proc. SPIE (Advances Optical Thin Films), **5250**, 384 (2004)
[56] J. Floriot, F. Lemarchand and M. Lequime, Applied Optics **45(7)**, (2006)
[57] M. Niraula, J. W. Yoon, and R. Magnusson, arXiv:1509.02974 (2015)
[58] V. Leonhardt and J.B. Camp, Applied Optics, **45(17)**, 4142 (2006)
[59] Ke-Xun Sun *et al*. Classical Quant. Grav. **22**, S287 (2005)
[60] M. Armano *et al*. Phys. Rev. Lett. 116 (23): 231101 (2016)
[61] M. Tinto, F.B. Estabrook, J.W. Armstrong, Phys. Rev. D, **65**, 082003 (2002)
[62] M. Tinto and N. Yu, Phys. Rev. D, **92**, 042002 (2015)
[63] G. Heinzel, 11th LISA Symposium, Zurich, Switzerland, (2016)
[64] T. Wilken *et al*. OSA Technical Digest, Paper AF2H.5 (2013)
[65] Space-qualified optical frequency combs
[66] C. L. Wainwright and T. A. Moore, Phys. Rev. D, **79(2):024022**, (2009)
[67] R. Stebbins *et al.* LISA: Response to Astro 2010 RFI for the Particle Astrophysics and Gravitation Panel, (2009)
[68] I. Thorpe and J. Livas, NASA's Gravitational-Wave Mission Concept Study for the GW Study Team, Physics of the Cosmos Program Analysis, Group Meeting Washington DC, August 14th, (2012)
[69] J. Li *et al*. Advances in Space Research **40(1)** 1 (2007)
[70] O. Jennrich *et al*, *NGO assessment study report* (*Yellow Book*) ⟨hal-00730260⟩ (2012)
[71] A. Zoellner *et al*. Optics Express, **25(21),** 25201 (2017)
[72] K. Sun, *et al.* AIP Conference Series, **873**, 515 (2006)
[73] S. Buchman and J. P. Turneaure, Rev. Sci. Instrum. **82**, 074502. (2011)
[74] D. B. DeBra, *Disturbance compensation system design*, APL Technical Digest, 12(22), 14 (1973)
[75] Staff of the Space Department *et al*. American Inst. of Aero. & Astro. Journal (AIAA), **11**, 637 (1974)
[76] C. W. F. Everitt *et al*. Phys. Rev. Lett. **106**, 221101 (2011)
[77] D. Gerardi *et al*. Rev. Sci. Instrum. **85**, 011301 (2014)
[78] J. Conklin *et al.* Proceedings of the AIAA/USU Conference on Small Satellites, SSC12-VI-8 (2012)
[79] S. Saraf *et al*. Classical Quant. Grav. **33**, 245004 (2016)
[80] NASA: Gravitational-Wave Mission Concept Study Final Report (2012)
[81] European Space Agency, NGO: Revealing a hidden Universe: opening a new chapter of discovery, Assessment Study Report (2011)
[82] NASA, Physics of the Cosmos
[83] U.A. Johann *et al*. AIP Conf. Proc., **873**, 304 (2006)
[84] A. J. Mullavey *et al*. Optics Express, **20(1)** 81 (2012)
[85] F Antonucci *et al.* arXiv:1012.5968v3 (2012)
[86] Committee for a Decadal Survey of Astronomy and Astrophysics (2010)
[87] B. Zink, *et al*, Phys. Rev. Lett. **96**, 161101 (2006)